# A Hypothesis about the Physical Nature of the Phenomenon of Life (A Contribution to the Discussion of the Paper by G.R. Ivanitskii 21st Century: What Is Life from the Perspective of Physics)

A. M. Smolovich*

*Kotel'nikov Institute of Radioengineering and Electronics, Russian Academy of Sciences, Moscow, 125009 Russia*
*e-mail: asmolovich@petersmol.ru*

**Abstract**—A macroscopic quantum state that is characterized by an energy gap in the electronic spectrum and occurs in organic structural components of the cell is hypothesized to provide a basis for the phenomenon of life. The width of the band gap has been estimated. The possibility of detecting this energy gap in an experimental study is discussed.

**Keywords:** phenomenon of life, Cooper pairs, energy gap, optical spectroscopy

In his article "21st Century: What Is Life from the Perspective of Physics," Ivanitskii [1] paid considerable attention to the problem of providing life with a definition that would make it possible to differentiate animate and inanimate matters by the presence or absence of a particular trait or a set of traits. The problem is very difficult. Hypotheses about the origin of life were additionally discussed in the article. The problems aroused great interest and prompted subsequent discussion. However, it is possible to use another approach to the problem specified in the title of the article [1]. An increasing number of researchers come to recognize that something very important remains unknown [2], while the structure has been determined almost to the atomic level in some primitive living organisms. What can we not know while understanding the structure in detail? One possible answer is that the state of the structure remains obscure. If life is considered as a special state that is characteristic of organic structural components of the cell and is still unknown in terms of its physical nature, than the phrase "what is life from the prospective of physics" can be understood as a question about the nature of life, that is, the physical mechanism that forms its basis. It seems natural in this case to find analogs among the phenomena studied in condensed matter physics. A similar approach has already been proposed earlier by a number of famous physicists.

London [3] assumed in a preface to his monograph that a quantum behavior similar to that in superconductivity and superfluidity can occur and to play a role in biological processes. According to London, certain interactions between macromolecules in biochemistry can be understood only as a result of a quantum mechanism that is characteristic of the total system. This provides the system with the characteristic stability of quantum states with the ability to change without involving dissipation processes. Little [4, 5] drew attention to London's idea and noted that the idea is absolutely new and important for understanding living systems. Little then assumed that superconductivity at temperatures close to room temperature can be reached in an organic polymer that is structurally similar to DNA.

Local superconductivity problems were considered more recently at the molecular level. Complex molecules with conjugated bonds have electrons in both inner ($\sigma$ electrons) and outer ($\pi$ electrons) orbitals [6]. The $\pi$ electrons can move along the total molecular $\sigma$ backbone (atomic nuclei with $\sigma$ electrons of the molecule), thus being collective delocalized electrons. In other words, the $\pi$ electrons are similar to free electrons in a conductor. Using the approach described in [7], Kresin [8] showed that a pair correlation similar to the formation of Cooper pairs in a superconductor is possible in a $\pi$-electron system. The association of electrons in a pair is due to their interaction with the $\sigma$ backbone, which plays a role similar to that of the crystal lattice in a superconductor. Complex molecules with conjugated bonds are components of biologically active substances. Kresin [8, 9] refers to a book by B. and A. Pullmans [10]. The book ends with the chapter *Electron Delocalization and the Processes of Life*. A conclusion reached in the chapter is that the presence of an electronic cloud in conjugated molecules can be considered as a main basis of life. Kresin assumed that a pair correlation of collectivized electrons, which results in a band gap arising in the spectrum, ensures stability similar to that observed in superconductive metals, and a long-range order due to electron–electron correlation is essential for understanding the mechanism of connection, excitation transfer in biologically active substances (see the section *Superconductivity and Physics of Complex Molecules* in [8]).

To clarify, a superconducting state does not mean that electrical current can be carried with zero resistance in the case of complex molecules. This is only one of the characteristic (but not in this case) properties of the superconducting state, and there are others. One of these is that an energy gap occurs in the density of states of electrons in the vicinity of the Fermi level; there is no allowed energy level within the gap. The gap width $2\Delta$ corresponds to the binding energy of a Cooper pair of electrons. The gap stabilizes the superconducting state because energy approximately equal to the gap width is necessary to spend to disrupt the state. The Bardeen–Cooper–Schrieffer theory predicts the gap width for conventional superconductors at $T = 0$: $2\Delta = 3.52 k_B T_c$, where $T_c$ is the superconducting transition temperature and $k_B$ is Boltzmann's constant (Eq. (45.18) in [11]). The energy gap sizes of high-temperature superconductors [12] and charge-density waves [13] differ from the above equation, but are similarly of the order of $2k_B T_a$, where $T_a$ is the temperature of the superconducting transition or Peierls junction, respectively.

My hypothesis is that a macroscopic quantum state of organic structures occurring in the cell underlies the phenomenon of life [30]. The state presumably differs from superconductivity, but its stability is similarly ensured by the presence of an energy gap in the electron spec- trum. The energy gap width is assumed to be similarly of the order of $2k_B T_V$, where $T_V$ is a certain temperature typical of living organisms, e.g., 300 K. With this value, the energy gap width is estimated to be approx-imately $5 \times 10^{-2}$ eV. One may attempt to detect the energy gap experimentally.

The following experimental techniques can be used to identify the energy gap. Tunneling spectroscopy is broadly used to measure the density of states and, in particular, to identify the superconducting and charge-density wave energy gaps [14]. The technique was applied to biological objects as well. As an example, single DNA molecules were studied by scanning tunneling spectroscopy [15–17]. However, a rather great scatter of measurements was reported in these studies. Conventional optical spectroscopy in various wavelength ranges of the electromagnetic spectrum is also used to identify the energy gap (see [18, 19] and references therein). The method is contactless and therefore appealing for studies of living objects. A spectral range from 600 to 2000 cm$^{-1}$ was used in the majority of studies with biological objects. The following optical spectroscopy techniques have been used to study bacteria and phages: Raman and surface-enhanced Raman spectroscopy [20–24], coherent anti-Stokes Raman spectroscopy [25], Fourier transform infrared spectroscopy [26], and time-resolved infrared spectroscopy [27, 28]. A spectroscopy study was carried out with single bacteria, which were immobilized in the confocal microscope field of view with so-called Raman tweezers [24]. The method seems to be preferable for identifying the energy gap in living structures.

The question of where and how to look for the energy gap is a matter of discussion. It is clear that primitive organisms should be chosen for the study. Simple biological objects, such as prokaryotes and viruses, are far more complex than the majority of objects examined in physical studies. Their spectra are also rather complex, complicating the identification of the energy gap. When bacteria are used in such studies, one can compare their spectra between normal and inactivated states, which would serve as markers of the living (V, vita) and dead (M, mort) states (designated as in [1]). The study will be analogous to comparing the spectra obtained for a superconductor at temperatures higher and lower than the superconducting transition point (see Fig. 2 in [18] and Fig. 1 in [19]). However, well-known processes related to vital activities take place in the living cell and will complicate the identification of the energy gap in the spectrum of a bacterium. The processes also contribute to the difference in spectra between viable and dead cells [23]. Spectra of individual cell organelles or individual biological molecules contained in the cell, such as protein molecules, may be expedient to use in a search for the energy gap.

Another variant is using the spectra of viruses to look for the energy gap. There is no consensus as to whether viruses are a form of life or organic structures that interact with living organisms. Viruses were termed "organisms at the edge of life" in [1]. Outside of the host cell, viruses do not display vital characteristics, such as metabolism, replication, etc. Viruses occurring within and outside the host cell can be used as V and M markers. Infection of bacteria with phages is suitable for such studies. A comparison is performed between phage DNA spectra obtained before phage DNA enters the cell and during its persistence within the cell in this case. The design takes into account that only phage DNA enters the bacterial cells upon its infection, while the protein envelope (capsid) of the phage remains outside of the membrane. In addition, when a bacterial cell is infected with phage DNA, its vital processes change as a result of infection. Phage DNA is directly involved in certain cell processes, which will change its spectrum and complicate the identification of the energy gap. One special problem is to separate the spectra of phages, bacteria, and the solutions where they occur. Special methods were designed for this purpose (see [29]) and successfully used, in particular, in spectroscopy studies of phages and bacteria. The methods make it possible to distinguish between bacterial and phage spectra and between spectra of different phage species [20].

To summarize, we discussed a hypothesis that considers the phenomenon of life as a macroscopic quantum state of organic structures contained in the cell; the state is characterized by an energy gap present in the electron spectrum. The possibility of experimentally identifying the gap was also considered.

ACKNOWLEDGMENTS

I am grateful to E.R. Lozovskaya, A.A. Sinchenko, S.V. Chekalin, A.V. Kalinin, and D.V. Klinov for helpful advice and discussion.